\documentclass[prl,twocolumn,superscriptaddress,amsmath,amssymb,amsfonts]{revtex4}

\usepackage[sort&compress]{natbib}

\newtheorem{theorem}{Theorem}
\newtheorem{lemma}{Lemma}
\newtheorem{proposition}{Proposition}

\newtheorem{corollary}{Corollary}
\newtheorem{definition}{Definition}

\newcommand{\Id}{\mbox{Id}}

\newcommand{\maxtensor}{\otimes_{max}}
\newcommand{\mintensor}{\otimes_{min}}

\renewcommand{\hat}{\widehat}

\newcommand{\ran}{\mbox{Ran}}
\newcommand{\f}{{\bf f}}
\renewcommand{\H}{{\bf H}}

\newcommand{\K}{{\bf K}}
\newcommand{\ostar}{\circledast}

\newcommand{\iso}{\simeq}

\newcommand{\fD}{{\frak D}}
\newcommand{\norm}[1]{\widetilde{#1}}
\newcommand{\beq}{\begin{equation}}
\newcommand{\eeq}{\end{equation}}

\begin{document}

\title{Teleportation in General Probabilistic Theories}

\author{Howard Barnum} \email{barnum@lanl.gov} \affiliation{CCS-3:
Information Sciences, MS B256, Los Alamos National Laboratory, Los
Alamos, NM 87545 USA}

\author{Jonathan Barrett} \email{jbarrett@perimeterinstitute.ca}
\affiliation{Perimeter Institute for Theoretical Physics, 31
Caroline Street N, Waterloo, Ontario N2L 2Y5, Canada}

\author{Matthew Leifer} \email{matt@mattleifer.info}
\affiliation{Institute for Quantum Computing, University of Waterloo,
Waterloo, Ontario, Canada}

\author{Alex Wilce} \email{wilce@susqu.edu} \affiliation{Department of
Mathematical Sciences, Susquehanna University, Selinsgrove, PA 17870
USA}

\begin{abstract}
In a previous paper, we showed that many important quantum
information-theoretic phenomena, including the no-cloning and
no-broadcasting theorems, are in fact generic in all non-classical
probabilistic theories. An exception is teleportation, which most
such theories do not support. In this paper, we
investigate which probabilistic theories, and more particularly,
which composite systems, {\em do} support a teleportation protocol.
We isolate a natural class of composite systems that we term {\em
regular}, and establish necessary and sufficient conditions for a
regular tripartite system to support a
conclusive, or post-selected, teleportation protocol.  We
also establish a sufficient condition for deterministic
teleportation that yields a large supply of theories, neither classical
nor quantum, that support such a protocol.
\end{abstract}

\maketitle

%\section{Introduction}
The standard quantum teleportation protocol \cite{Bennett et al}
allows two parties, Alice and Bob, to transmit an unknown quantum
state from Alice's site to Bob's; in compliance with the no-cloning
theorem, Alice's copy is destroyed in the process. The protocol
assumes that Alice and Bob have access to the two wings, $A$ and
$B$, of a bipartite system $A \otimes B$ in a maximally entangled
state, which serves as a kind of quantum channel. The state to be
teleported belongs to an auxiliary system $A'$ at Alice's site,
which is coupled to her half of the shared system. Alice measures an
observable corresponding to the Bell basis on the combined system
$A' \otimes A$. Depending upon the result, she instructs Bob (via
purely classical signaling) to perform a particular unitary
correction on his wing, $B$, of the shared $A \otimes B$ system.
With certainty, Alice now knows that the state of Bob's system is
identical to the state (whatever it was) of her ancillary system
$A'$.

The possibility of teleportation is surprising, in view of the
no-cloning and no-broadcasting theorems, which prohibit the copying
of quantum information. In a previous paper [3], we have shown that
both no-cloning and no-broadcasting theorems are in fact quite
generic features of essentially any {\em non-classical}
probabilistic theory, and not specifically quantum at all. However,
as pointed out in \cite{Barrett, BBLW06}, most such theories {\em do
not} allow for teleportation.  Classical theories, however, do.
The possibility of teleportation can
thus be regarded, in some very rough qualitative sense, as a measure
of the relative {\em classicality} (or at any rate, {\em tameness})
of quantum theory.
%, within the spectrum of probabilistic theories
%that are mathematically possible.

In this note, we make some precise statements about {\em which}
probabilistic theories---and more particularly, which tripartite
systems---admit teleportation. For simplicity, consider the case in
which the three component systems, $A', A$ and $B$, in the protocol
described above are identical. Then an obvious necessary condition
for the protocol to succeed is that the cone of unnormalized states
in $A$ be isomorphic to the {\em dual} cone of unnormalized {\em
effects} in $A^{\ast}$---a strong condition that is nevertheless
satisfied by both quantum and classical systems. As we shall see,
this is sufficient to ground conclusive (or one-outcome
post-selected) teleportation. To obtain deterministic teleportation
appears to be more difficult; however, where the state space has
sufficient symmetry, a sort of deterministic teleportation can
always be achieved with respect to a possibly continuously-indexed
observable. Specializing to the case in which the state space is
symmetric under the action of a finite group, we obtain a wealth of
examples of state spaces that are neither classical nor
quantum-mechanical, but nevertheless support a
genuine deterministic teleportation protocol. \\%%Vinberg etc.?

%\section{Background}
\noindent{\bf 1. Probabilistic Models} This section assembles the
necessary machinery of generalized probability theory---
essentially, the convex sets framework deriving from the work of
Mackey \cite{Mackey} and subsequently refined by many authors,
notably Davies and Lewis \cite{DaviesLewis}, C. M. Edwards
\cite{Edwards} and Ludwig \cite{Ludwig}. We use more or less the
same notation as in \cite{BBLW06, BBLW07}; as in the latter, in this
paper we consider only probabilistic models having
finite-dimensional state spaces. \\

\noindent \emph{Abstract State Spaces} We model a physical system by
an ordered vector space $A$ with a (closed, pointed, generating)
positive cone $A_{+}$, which we regard as consisting of
un-normalized ``states". We also posit a distinguished order unit,
that is, a linear functional $u_A$ that is {\em strictly} positive
on non-zero positive elements of $A$; this defines a compact convex
set $\Omega_{A} = u_{A}^{-1}(1)$ of {\em normalized} states. We
shall call an ordered linear space, equipped with such a
functional---more formally: a pair $(A,u_{A})$---an {\em abstract
state space}. If $(A,u_A)$ and $(B,u_B)$ are abstract state spaces,
we write $A \leq B$ to indicate that (i) $A$ is a subspace of $B$;
(ii) $A_+ \subseteq B_+$; and (iii) $u_A$ is the restriction of
$u_B$ to $A_+$. Similarly, $A \simeq B$, read ``$A$ is isomorphic to
$B$'', means that there exists an invertible, positive linear
mapping $A \rightarrow B$, with a positive inverse, and taking the
order unit of $A$ to that of $B$.  Equivalently, such a mapping
takes $A$'s normalized state space $\Omega_{A}$ bijectively (and
affinely) onto $B$'s normalized state space $\Omega_{B}$.  We refer
to an isomorphism $A \rightarrow A$ as a {\em symmetry} of $A$.  A
positive linear mapping with positive inverse, but that does not
necessarily preserve the order unit, we refer to as an {\em order
isomorphism} between $A$ and $B$, and we say they are
order-isomorphic if such a map exists.

By way of illustration, discrete classical probability theory
concerns the case in which $A$ is the space ${\Bbb R}^{E}$ of all
real-valued functions $\alpha$ on a finite set $E$ of measurement
outcomes, in the natural point-wise ordering. The order unit is the
functional $u_{A}(\alpha) := \sum_{x \in E} \alpha(x)$, hence the
normalized state space $\Omega_{A}$ consists of all probability
weights on $E$. In elementary quantum probability theory, $A$ is the
space of Hermitian operators on a complex Hilbert space $\H$,
ordered in the usual way; the order unit is the trace, so that
$\Omega_{A}$ is the set of density operators.

Physical events (e.g., measurement outcomes) associated with an
abstract state space $A$ are represented by {\em effects}, that is,
positive linear functionals $f \in A^{\ast}$ with $f(\alpha) \leq 1$
for all $\alpha \in \Omega_A$, or, equivalently, $f \leq u_{A}$. The
understanding is that $f(\alpha)$ represents the {\em probability}
that the event in question will occur when the system's state is
$\alpha$. As indicated above, we wish to restrict our attention here to
cases in which the space $A$ is finite-dimensional. Thus we may
identify $A$ with $A^{\ast \ast}$, so that, for $\alpha \in A$ and
$f \in A^{\ast}$, we may write $f(\alpha)$ as $\alpha(f)$ whenever
it suits us.  In the sequel, we shall continue always to denote
states by lower case Greek letters, and effects, by lower case Roman
letters.

It is helpful to note that the set $\Omega_{A}$ of normalized states
actually determines both the ordered space $A$ and the order-unit
$u_{A}$: one can take $A$ to be the dual of the space of affine
real-valued functionals on $\Omega_{A}$, ordered by the cone of
non-negative affine functionals; $u_{A}$ is simply the constant
affine functional on $\Omega_{A}$ with value $1$. When describing a
particular abstract state space, it is often easiest simply to
specify the convex set $\Omega_{A}$. When we wish to begin with a
convex set $\Omega$ and reconstruct $A$ in this way, we write $A =
A(\Omega)$.   %%Or $V(\Omega)$?

Note that the point-wise ordering of functionals in $A^{\ast}$ on
$\Omega_{A}$ is exactly the usual dual ordering. There is a natural
norm on $A^{\ast}$, namely the supremum norm $\|f\| = \sup_{\alpha
\in \Omega_{A}} |f(\alpha)|$; this gives rise in turn to a norm on
$A$, called the {\em base norm}, with respect to which $\|\alpha\| =
u_{A}(\alpha)$ for $\alpha \in A_{+}$. In particular, every
normalized state has norm $1$, and conversely, a positive element of
$A$ having norm $1$ is a normalized state (so that the two meanings
of ``normalized" coincide). In the sequel, we shall write
$\norm{\alpha}$ for the normalized version of a positive weight
$\alpha \in A_+$, i.e.,
\[\norm{\alpha} := \frac{\alpha}{\|\alpha\|} =
\frac{\alpha}{u_{A}(\alpha)}\] It will be convenient to stipulate
that $\norm{0} = 0$.       \\

\noindent\emph{Observables} Let $(X,{\cal B})$ be a measurable
space: an {\em $X$-valued observable} on a state space $A$ is a
weakly countably additive vector measure $F : {\cal B} \rightarrow
A^{\ast}$ with $F(X) = u$. This guarantees that if $\alpha \in
\Omega$, $B \mapsto F(B)(\alpha)$ is a (finitely additive)
probability measure on ${\cal B}$. If $\mu$ is a given measure on
$(X,{\frak B})$, we shall call $f : X \rightarrow A^{\ast}$ a {\em
density} for $F$ with respect to $\mu$ iff, for every $\alpha \in
\Omega$ and every set $B \in {\cal B}$,
\[\int_{B} f(x)(\alpha) d\mu(x) = F(B)(\alpha).\]

In the simplest case, where $X$ is a finite set and $\mu$ is the
counting measure, an $X$-valued observable amounts to a list
$(f_1,...,f_n)$ of effects with $\sum_i f_i = u$. In the sequel,
when we speak of {\em an observable}, without specifying the value
space, this is what we have in mind.  \\

\noindent\emph{Processes and Dynamics} We represent physical {\em
processes} involving an initial system with state space $A$ and a
final system with state space $B$ by positive linear mappings $\phi
: A \rightarrow B$ having the property that $\|\phi(\alpha)\| \leq
1$ for all $\alpha \in \Omega_{A}$, which is just to say that $\phi$
is norm contractive, or, equivalently, that $\|\phi\| \leq 1$. In
this case, we understand that $\|\phi(\alpha)\| = u(\phi(\alpha))$
represents the probability that the process occurs when the input
is $\alpha$; indeed, we can regard the effect $u \circ \phi \in
A^{\ast}$ as recording precisely this occurrence. Thus, a family
$\{\phi_{i} | i \in I\}$ of positive linear mappings with
$\|\phi_i\| \leq 1$ for all $i$ and $\sum_{i} \|\phi_i\| = 1$,
represents a family of physical processes {\em one} of which is
bound to occur. (Such a family is a (discrete) {\em instrument} in
the sense of \cite{DaviesLewis, HolevoJMP}.)

In many cases, one wants to impose some further constraint on the
possible dynamics of a system represented by an abstract state space
$A$. By a {\em dynamical semigroup} for $A$, we mean a closed, convex set
${\frak D}_{A}$ of norm-contractive positive linear mappings $\tau :
A \rightarrow A$, closed under composition and containing the
identity mapping $\Id_A$. We understand ${\frak D}_{A}$ as representing the
set of all physically possible processes on $A$. (Here,
``physically'' refers to the use of this framework for abstractly
formulating possible physical theories;  the framework could also
have other applications, so the terminology ``operationally
possible'' might be more accurate.  With this caveat, however, we
will stick with ``physically.'')   A state space equipped with a
distinguished dynamical semigroup, we call a {\em dynamical model}. Note
that any abstract state space can be regarded as a dynamical model
if we take ${\frak D}_{A}$ to be (by default) the set of all
norm-contractive positive linear mappings $A \rightarrow A$. {\em In
the balance of this paper, we take it as a standing assumption,
relaxed only where explicitly noted,
that this is the case. } \\

\noindent\emph{Self-Duality and Weak Self-Duality} In both classical
and quantum settings, $A$ carries a natural inner product with
respect to which there is a {\em canonical} order-isomorphism $A
\simeq A^{\ast}$. Indeed, in both classical and quantum cases, the
positive cone is {\em self dual}, in that
\[A_{+} = A^{+} := \{
\alpha \in V | \forall \beta \in V_{+} \langle \alpha, \beta \rangle
\geq 0\}.\]
This property is a very special one, not shared by most
abstract state spaces. For an example, let $A$ be three-dimensional,
with $\Omega_{A}$ a square. For each side of the square, there is an
effect taking the value $1$ along that side, with the effects
corresponding to opposite sides summing to $1$. The dual cone thus
also has a square cross-section, so that the cones $A_{+}$ and
$A^{\ast}_{+}$ are isomorphic. Nevertheless, $A_{+}$ is not
self-dual, as $A^{+}$ is the image of $A_{+}$ under a rotation by
$\pi/4$.

In this paper, we shall call a finite-dimensional ordered space {\em
weakly self-dual} iff, as in the example above, there exists an
order isomorphism (that is, a bijective, positive linear mapping
with positive inverse)  $\phi : V \simeq V^{\ast}$. This is a far
less stringent condition than self-duality. A classical result of
Vinberg \cite{Vinberg} and Koecher \cite{Koecher} shows that any
finite-dimensional self-dual cone that is {\em homogeneous}, in the
sense that any interior point can be mapped to any other by an
affine symmetry (automorphism) of the cone, and {\em irreducible} in
the sense that the cone is not a direct sum of simpler cones, is
either the cone of positive self-adjoint elements of some full
matrix $\ast$-algebra over the reals, complexes or quaternions, or
is the cone generated by a ball-shaped base, or is the set of
positive self-adjoint $3 \times 3$ matrices over the octonions.

Thus, self-duality, plus irreducibility and homogeneity, brings us
within hailing distance of Hilbert space quantum mechanics. One
might hope to motivate these conditions in operational terms. In
this paper, we make some progress in this direction by identifying
{\em weak} self-duality of a system as a necessary
condition for a composite of
three copies of the system to support conclusive (probabilistic)
teleportation, and a condition not much stronger than
homogeneity on the space of
normalized states of the system to be teleported, as sufficient for
the existence of a tripartite model permitting
deterministic teleportation.\\

\noindent{\bf 2. Composite Systems} In order to discuss
teleportation protocols, it is important to consider composite
systems having, at a minimum, three components: one corresponding to
the sender (``Alice"), another to the receiver (``Bob"), and a
third, accessible to the sender but entangled with the receiver, to
serve as a channel across which the sender's state can be
teleported. In this section, we review the account of bipartite
state spaces given in \cite{BBLW06}, and extend it to cover systems
having three or more components. In doing so, we identify a
non-trivial condition on such composites, which we term {\em
regularity}, that will play an important role in our discussion of
teleportation protocols in the sequel.

In order to maintain the flow of discussion, the proofs of several
results from this section have been placed in a brief appendix.\\

\noindent\emph {Bipartite Systems} It will be convenient, in what
follows, to identify the algebraic tensor product, $A \otimes B$, of
two vector spaces $A$ and $B$ with the space of all bilinear forms
on $A^{\ast} \times B^{\ast}$. In particular, if $\alpha \in A$ and
$\beta \in B$, we identify the pure tensor $\alpha \otimes \beta$
with the bilinear form defined by
\[(\alpha \otimes \beta)(a,b) = a(\alpha)b(\beta)\]
for all $a \in A^{\ast}$ and $b \in B^{\ast}$. If $A$ and $B$ are
{\em ordered} vector spaces, we call a form $\omega \in A \otimes B$
{\em positive} iff $\omega(a,b) \geq 0$ for all positive functionals
$a \in A^{\ast}$ and $b \in B^{\ast}$. Note that if $\alpha \in A$
and $\beta \in B$ are positive, then $\alpha \otimes \beta$ is a
positive form. Note, too, that the set of positive forms is a cone
in $A \otimes B$.

\begin{definition} The {\em maximal tensor product} of ordered vector spaces $A$
and $B$, denoted $A \maxtensor B$, is $A \otimes B$, equipped with
the cone of all positive forms. Their {\em minimal tensor product},
denoted $A \mintensor B$, is $A \otimes B$ equipped with the cone of
all positive linear combinations of pure tensors.
\end{definition}

The maximal and minimal tensor products are exactly the injective
and projective tensor products discussed by Wittstock in
\cite{Wittstock}; see also \cite{Ellis, Namioka-Phelps}. It is not
difficult to show that, in our present finite-dimensional setting,
$(A \maxtensor B)^{\ast} = A^{\ast} \mintensor B^{\ast}$ and $(A
\mintensor B)^{\ast} = A^{\ast} \maxtensor B^{\ast}$.
% We shall make
%frequent use of these dualities.
%Note that, as a
%consequence, we have $A \maxtensor B = (A^{\ast} \mintensor
%B^{\ast})^{\ast}$ and $A \mintensor B = (A^{\ast} \maxtensor
%B^{\ast})^{\ast}$.

If $(A,u_{A})$ and $(B,u_{B})$ are abstract state spaces
representing two physical systems, then subject to a plausible
no-signaling condition and a ``local observability'' assumption
guaranteeing that the correlations between local observables
determine the gloabl state
(see \cite{BBLW06}), the largest sensible
model for a bipartite system having physically separated components
modeled by $A$ and $B$ is $A \maxtensor B$, with order unit given by
$u^{AB} = u_{A} \otimes u_{B}$. Accordingly, we model a {\em composite
system} with components $(A,u_A)$ and $(B,u_B)$ by the algebraic
tensor product of $A$ and $B$, ordered by {\em any} cone lying
between the maximal and minimal tensor cones, and with order unit
$u_{AB} = u_{A} \otimes u_{B}$. We shall write $AB$, generically,
for such a state space, denoting the convex set $u_{AB}^{-1}(1)$ of
normalized states by $\Omega_{AB}$.

It will be important, below, to remember that all states, in whatever
cone we use, can be represented as linear combinations of pure
product states, as these span $A \otimes B$. Unless the sets of
normalized states for $A$ or $B$ are simplices---that is, unless
one system at least is classical---the minimal and maximal tensor
products are quite different, with the latter containing many more
normalized states than the former. These additional states we term
{\em entangled}; states in $A \mintensor B$, we term {\em
separable}.\\

\noindent{\em Marginal and Conditional States} Every state $\omega$
in a bipartite system $AB$ has natural {\em marginal states}
$\omega^A \in A$ and $\omega^{B} \in B$, given respectively by
\[\omega^{A}(a) = \omega(a \otimes u_{B}) \ \text{and} \ \omega^{B}(b) =
\omega(u_{A} \otimes b)\] for all $a \in A^{\ast}$ and $b \in
B^{\ast}$. We also have un-normalized conditional states, given by
\[\omega^{B}_{a}(b) = \omega(a,b) = \omega^{A}_{b}(a)\]
and their normalized versions,
\[\norm{\omega}^{B}_{a}(b) = \frac{\omega(a,b)}{\omega^{A}(a)} \ \text{and~} \omega^{A}_{b}(a) =
\frac{\omega(a,b)}{\omega^{B}(b)}\] if the marginal states are
non-zero, and set equal to $0$ otherwise, so that the expected
identities $\omega(a,b) = \omega^{A}_{b}(a)\omega^{B}(b) =
\omega^{A}(a)\omega^{B}_{a}(b)$ hold. Using these, it is not
difficult to show that, just as in quantum theory, the marginals of
an entangled state are necessarily mixed, while those of an
unentangled pure state are necessarily pure.\\

\noindent{\em Dynamically Admissible Composites} It is reasonable to
suppose that, if $\tau_{A} \in {\frak D}_{A}$ and $\tau_B \in {\frak
D}_{B}$ are physically admissible processes on $A$ and $B$,
respectively, then, for any state $\omega$ on a composite system
$AB$, \[(\tau_A \otimes \tau_B)(\omega) : a,b \mapsto
\omega(\tau_{A}^{\ast} a, \tau^{\ast}_{B} b)\] is a state of $AB$.
When this is the case, let us say that the composite system $AB$ is
{\em dynamically admissible}. Equivalently, $AB$ is dynamically
admissible iff for all $\tau_A \in {\frak D}_A, \tau_B \in {\frak
D}_B$, $AB_{+}$ is stable under $\tau_{A} \otimes \tau_{B}$ acting
on $A \otimes B$. Note that both minimal and maximal tensor products
are stable under any pure tensor of positive operators, so these are
dynamically admissible regardless of the dynamics.

Where ${\frak D}_{A}$ and ${\frak D}_{B}$ -- as per our standing
assumption -- comprise {\em all} norm-contractive positive mappings $A
\rightarrow A$ and $B \rightarrow B$, respectively, $AB$ is
dynamically admissible iff its positive cone $AB_+$ is stable under
$\tau_1 \otimes \tau_2$ for {\em all} positive mappings $\tau_1 : A
\rightarrow A$ and $\tau_2 : B \rightarrow B$.  Although the minimal
and maximal tensor products $A \mintensor B$ and $A \maxtensor B$ both
enjoy this property, it is highly non-trivial. Indeed, if $A = {\cal
  B}_{h}(\H)$ and $B = {\cal B}_{h}(\K)$, the spaces of self-adjoint
operators on Hilbert spaces $\H$ and $\K$, and $AB = {\cal B}_{h}(\H
\otimes \K)$, the usual quantum-mechanical composite state space,
then the cone $AB_{+}$ is stable only under products of {\em
completely} positive mappings. However, this difficulty is easily
met: one need only define a composite of two dynamical models
$(A,{\frak D}_{A})$ and $(B,{\frak D}_{B})$ to be a model $(AB,
{\frak D}_{AB})$ where $AB$ is a dynamically admissible composite of
$A$ and $B$, and ${\frak D}_{AB}$ is a semigroup of norm-contractive
positive mappings $AB \rightarrow AB$ containing all products
$\tau_{A} \otimes \tau_{B}$ where $\tau_{A} \in {\frak D}_{B}$ and
$\tau_{B} \in {\frak D}_{B}$. In the balance of this paper, results
will be formulated for composites of state spaces, rather than of
dynamical models; however, these can easily be modified to
accommodate the latter. \\

\noindent\emph{Bipartite states and effects as operators} Elements
of the tensor product $A \otimes B$ and of its dual $(A \otimes
B)^{\ast}$ can be regarded as operators $A^{\ast} \rightarrow B$ and
$A \rightarrow B^{\ast}$, respectively. Indeed, every $f \in (A
\otimes B)^{\ast}$ induces a linear mapping $\hat{f} : A \rightarrow
B^{\ast}$, uniquely defined by the condition that
\[\hat{f}(\alpha)(\beta) = f(\alpha \otimes \beta).\]
The mapping $f \mapsto \hat{f}$ is a linear isomorphism. Note also
that, if $f$ is positive, then so is $\hat{f}$ (though not
conversely, unless we use the maximal tensor product). Similarly,
any $\omega \in A \otimes B$ induces a linear mapping $\hat{\omega}
: A^{\ast} \rightarrow B$, uniquely defined by the condition that
\[\hat{\omega}(f)(g) = (f \otimes g)(\omega)\]
for all $f, g \in V^{\ast}$. Again, the mapping $\omega \mapsto
\hat{\omega}$ is a linear isomorphism. Also, since elements of the
maximal tensor product $A \maxtensor B$ are precisely those
corresponding to positive bilinear forms, $\hat{\omega}$ will be a
positive operator, regardless of which tensor product we use. In the
special case in which $\omega$ is a pure tensor, say $\omega = \beta
\otimes \gamma$, we have
\[\widehat{(\beta \otimes \gamma)}(f) = f(\beta) \gamma.\]
In the sequel, we shall write $\widehat{AB}$ for the set of
operators $\hat{\omega}$ corresponding to $\omega \in AB$, ordered
by the cone of operators $\hat{\omega}$ with $\omega \in AB_{+}$.
For example, $\widehat{A \maxtensor B}$ is simply the space ${\cal
L}(A,B)$, ordered by the cone of positive operators.

Note that the operator $\hat{\omega}$ corresponding to a {\em
normalized} state in $AB$ has the property that $\hat{\omega}(u)(u)
= 1$, i.e., $\hat{\omega}(u)$ is a state. Conversely, given a
positive linear mapping $\phi : A^{\ast} \rightarrow B$ with the
property that $\phi(u_{A})$ is a state, the bilinear form
$\omega(a,b) := \phi(a)(b)$ defines an element of the maximal tensor
product, with $\phi = \hat{\omega}$. It is useful to note
(\cite{Ellis}, Equation 16) that any positive operator $\phi :
A^{\ast} \rightarrow B$ has operator norm (induced by the above-defined
order-unit and base norms on $A^*$ and $B$) given by
\[\|\phi\| = \|\phi(u)\|_{B}\]  where $\| \ \cdot \ \|_{B}$ denotes the base-norm on
$B$; hence, bipartite states correspond exactly to positive
operators of norm $1$.    %%%Need to stress that \omega(f) is the un-normalized cond. state! Also that physically invertible means symmetry.

Similarly, if $f$ is a bipartite effect in $A^{\ast} \maxtensor
A^{\ast}$, then the mapping $\hat{f} : A \rightarrow A^{\ast}$ takes
any state $\alpha$ to the effect $\hat{f}(\alpha)(\beta) = f(\alpha
\otimes \beta)$. Evidently, this is no greater than unity on
$\Omega$, so we have $\hat{f}(\alpha) \leq u$ for all $\alpha \in
\Omega$; conversely, any such positive
mapping defines a bipartite effect.\\

%\subsection{Multi-partite Systems}
\noindent\emph{Multi-partite Systems} Up to a point, the foregoing
considerations readily extend to composite systems involving more
than two components. Suppose $(A_1,u_1),...,(A_n,u_n)$ are abstract
state spaces. As above, call an $n$-linear form on $A_1^{\ast}
\times \cdots \times A_{n}^{\ast}$ {\em positive} iff it takes
non-negative values on all $n$-tuples $f = (f_1,...,f_n)$ of
positive functionals $f_{i} \in A_{i}^{\ast}$. Given states
$\alpha_i \in {A_{i}}_{+}$ for $i = 1,...,n$, the product state
$\alpha_1 \otimes \cdots \otimes \alpha_n$, defined by $(\otimes_i
\alpha_i)(f) = \Pi_i \alpha_i(f_i)$, is obviously positive in this
sense.

\begin{definition} A {\em composite} of state spaces $(A_i, u_i)$,
$i = 1,...,n$, is any space $A$ of $n$-linear forms on $A_1^{\ast}
\cdots A_{n}^{\ast}$, ordered by any cone of positive forms
containing all product states, and with with order-unit given by $u
= u_1 \otimes \cdots \otimes u_n$.
\end{definition}

This is equivalent to saying that $A$ contains all product states,
and $A^{\ast}$ contains all product effects. Examples of composites
of, say, three spaces $A, B$ and $C$ would include $A \maxtensor B
\maxtensor C$, $A \mintensor B \mintensor C$, and mixed composites
such as $A \mintensor (B \maxtensor C)$. Extending the terminology
of the previous section, we shall call a composite $A$ of state
spaces $(A_i, u_i)$ {\em dynamically admissible} iff $A_{+}$ is
stable under mappings of the form $\bigotimes_{i} \tau_i$ where
$\tau_i : A_i \rightarrow A_i$ are arbitrary positive mappings. A
product of dynamical models $(A_i, {\frak D}_i)$ is  a dynamical
model $(A,{\frak D})$ where $A$ is a dynamically admissible model of
$A_1,...,A_n$ and ${\frak D}$ is a dynamical semigroup that includes
all products of mappings $\tau_i \in {\frak D}_i$.   \\

\noindent\emph{Regular composites} Suppose now that $A$ is a
composite of $A_1,...,A_n$, and that $J \subseteq \{1,....,n\}$.
Given a list of positive linear functionals $f = (f_i) \in \Pi_{i
\in I \setminus J} A_{i}^{\ast}$ and a state $\omega \in A_{+}$, we
may define a $|J|$-linear form $\omega^{J}_f$ on $\Pi_{j \in J}
A_{j}^{\ast}$ by setting
\[\omega^{J}_{f}(g) = \omega(f \otimes g),\] where $(f \otimes
g)_{i}$ is $g_i$ if $i \in J$ and $f_i$ otherwise. We refer to
$\omega^{J}_{f}$ as a {\em partially evaluated} state. The set of
such partially-evaluated states $\omega^{J}_{f}$ generates a cone in
$\bigotimes_{j \in J} A_{j}$; together with the order unit
$\otimes_{j \in J}~ u_j$, this defines an abstract state space
$A^{J}$, which we call the {\em $J$-partial sub-system}, and which
we take to represent the subsystem corresponding to the set of
elementary systems $A_{j}$ with  $j \in J$.

In the simplest cases, we should expect that that a composite of
``elementary" systems $A_1,...,A_n$ can equally be regarded as a
composite of complex sub-systems $A^{J}$ obtained through an
arbitrary coarse-graining of the index set $I = \{1,...,n\}$. This
suggests the following

\begin{definition} A composite $A$ of state spaces $A_1,...,A_n$ is
{\em regular} iff, for all partitions $\{J_1,...,J_k\}$ of
$\{1,...,n\}$, $A$ is a composite, in the sense of Definition 1, of
the partial systems $A^{J_1},...,A^{J_k}$.\end{definition}

Equivalently, $A$ is a regular composite of $A_1,...,A_n$ iff for
all partitions $J_1,..,J_k$ of $\{1,..,n\}$, and for all sequences
of states $\mu_k \in A^{J_k}$, the product state $\bigotimes_{k}
\mu_k$ belongs to $A$, and for all sequences of effects $f_k \in
(A^{J_k})^{\ast}$, the product effect $\bigotimes_{k} f_{k}$ belongs
to $A^{\ast}$.

We regard regularity as an eminently reasonable restriction on a
model of a composite physical system, at least in cases in which the
components retain their separate identities (so that the systems are
``separated"). As we shall see in the sequel, regularity is
sufficient to ground a weak analogue of a teleportation protocol,
which we call {\em remote evaluation}. In the balance of this
section, we collect some examples of regular composites, and adduce
some technical results concerning the notion of regularity.

As a matter of notational convenience, we'll write $ABC$ for a
composite of three systems $A$, $B$ and $C$, denoting by $AB$, $BC$,
and $AC$ the three bipartite subsystems. In this case, the condition
that $ABC$ be regular amounts to requiring that
\[AB \mintensor C \leq ABC  \leq AB \maxtensor C\] and similarly
$A$ and $BC$ and for $AC$ and $B$. Equivalently, we require that
\[AB \mintensor C \leq ABC \ \text{and} \ (AB)^{\ast} \mintensor
C^{\ast} \leq (ABC)^{\ast}.\]

As an example, let us show that the mixed tensor product
\[A \mintensor (B \maxtensor C)\] is a regular composite of $A, B$
and $C$. The only interesting coarse-grainings here are
$\{\{A,B\},\{C\}\}$ and $\{\{A,C\},\{B\}\}$.  To analyze
the first of these, suppose that $\omega =
\sum_{i} t_i \alpha_i \otimes \mu_i$ where $\alpha_i \in A_{+}$ and
$\mu_i \in (B \maxtensor C)_{+}$. Then for all $c \in C^{\ast}$,
\[\omega^{AB}_{c} = \sum_i t_i \alpha_i \otimes \hat{\mu}_i(c),\] a
positive linear combination of positive elements of $A$ and $B$;
hence, $\omega^{AB}_{c} \in (A \mintensor B)_{+}$, so $AB = A
\mintensor B$. It follows that, if $\gamma \in C_{+}$, we have
\begin{eqnarray*}
\omega^{AB}_{c} \otimes
\gamma \in (A \mintensor B \mintensor C)_{+} \\
\leq ((A \mintensor B)
\maxtensor C)_{+}
= (AB \maxtensor C)_+ \;.
\end{eqnarray*}
A similar argument applies to the bipartition $\{\{A,C\},B\}$.

 In the next section (see Corollary 1), we'll
show that $A \maxtensor (B \mintensor C)$ is also regular. An
example of a non-regular composite is \[(A \mintensor A) \maxtensor
(A \mintensor A)\] where $A$ is weakly self-dual. This follows from
considerations involving entanglement swapping, as discussed in
section 6; we postpone further discussion of this
example until then.\\

The following lemma collects a number of facts about composites and
regular composites that will be used freely---and often tacitly---
in the sequel. (For a proof, see the appendix.)

\begin{lemma} Let $A$ be a composite of systems $A_1,...,A_n$. Then
\begin{itemize}
\item[(a)] If $K \subseteq J \subseteq \{1,...,n\}$, then
$(A^J)^K = A^K$.
\item[(b)] If $A$ is regular, then $(A^J)_+ = \{ \omega^J_{u}
| \omega \in A_+\}$.
\item[(c)] If $A$ is regular, so is $A^J$ for every $J
\subseteq \{1,...,n\}$.
\end{itemize}
\end{lemma}

\noindent\emph{Probabilistic Theories} Roughly, by a {\em
probabilistic theory}, we mean a class ${\cal C}$ of probabilistic
models---that is, abstract state-spaces---closed under some
construction or constructions whereby systems can be composed.
Examples would include the class of all classical systems (i.e.,
systems with simplicial state spaces), the class of all quantum
systems with the usual quantum-mechanical state space, the class
obtained by forming the maximal tensor products of quantum systems,
the convexified version of Spekkens' ``toy theory" \cite{Spekkens},
etc. In principle, this idea might be given a precise
category-theoretic formulation (something we expect to pursue in a
subsequent paper); here, we content ourselves with a more informal
treatment.

Consider a class ${\cal C}$ of state spaces equipped with a specific
coupling $A, B \mapsto A \circledast B$, where $A \circledast B$ is
a composite of $A$ and $B$. We shall call $\circledast$ {\em
associative} if for all $A, B, C \in {\cal C}$, $A
\circledast (B \circledast C) \simeq (A \circledast B) \circledast
C)$ under the obvious association mapping (defined on product states
by $\alpha \otimes (\beta \otimes \gamma) \mapsto (\alpha \otimes
\beta) \otimes \gamma$. The straightforward but tedious proof of
the following can be found in the appendix:

\begin{proposition} If $\circledast$ is associative, then for
all $A_1,...,A_n \in {\cal C}$, $A_1 \circledast \cdots \circledast
A_n$ is a regular composite of $A_1,...,A_n$.
\end{proposition}

It follows that composites constructed using only the maximal, or
only the minimal, tensor product are regular, as are composite
quantum systems. For later purposes, if ${\cal C}$ is a class of
abstract state spaces closed under an associative coupling $\ostar$
preserving isomorphism, we shall call the pair $({\cal C},\ostar)$ a
{\em monoidal theory}.  (By {\em preserving isomorphism}, we mean
that if $A \iso B$ and $C \iso D$, then $(A \ostar C) \iso (B \ostar
D)$.) It is by no means obvious that every sensible theory must be
monoidal, however -- for instance, we may wish to consider theories
in which one can form tripartite systems of the form $A \mintensor
(B \maxtensor C)$, in which there is maximal entanglement between
$B$ and $C$, but no entanglement at all between $A$ and either $B$
or $C$.  There is certainly precedent for such mixed tensorial
constructions, e.g., in Hardy's causaloid framework for quantum
gravity \cite{Hardy}. On the other hand, considerations involving
entanglement swapping, as spelled out in section 5, place some
nontrivial restrictions on non-monoidal theories.
\\

%We shall call a probabilistic theory {\em regular} iff all composite
%systems that it allows are regular composites. By Proposition 1, any
%monoidal category of abstract state spaces is regular. \\

{\em Remark:} In the interest of clarity, it will sometimes be
helpful in the sequel to adorn an element of a factor in a tensor
product with a superscript indicating to which factor it belongs,
writing, for instance, $\alpha \otimes \beta$ or $\alpha^A \otimes
\beta^B$ for product states in $A \otimes B$, or $f^{AB}$ for an
arbitrary bipartite effect in $(A \otimes B)^{\ast}$. On occasion,
both ornamented and unornamented forms---e.g., $\omega$ and
$\omega^{AB}$---may occur in the same calculation; when they do,
they refer to the same object.   \\

\noindent{\bf 3. Conclusive Teleportation}
%Suppose that Alice and
%Bob have access, respectively, to the partial systems $A = A_1 A_2$
%and $B$ of a regular tripartite system $AB = A_1 A_2 B$. Let $f$ be
%an effect on the bipartite system $A = A_1 A_2$, and let $\omega$ be
%a state on the bipartite system $A_2 B$. As discussed above, these
%correspond, respectively, to a positive operator $\hat{f} : A_1
%\rightarrow A_{2}^{\ast}$ and a positive operator $\hat{\omega} :
%A_{2}^{\ast} \rightarrow B$. If $\alpha$ is a state in $A_1$ and $b$
%is an effect in $B^{\ast}$, then, by definition of regularity,
%$\alpha \otimes \omega$ is a state in $A_1 A_2 B$ and $f \otimes b$
%is an effect on $ABC$. In particular, $(f \otimes b) (\alpha \otimes
%\omega) \geq 0$.
Suppose $ABC$ is a composite of state spaces $A$, $B$ and $C$. If
$f$ is an effect on $AB$ and $\omega$ is a state in $BC$, then we
have positive linear mappings $\hat{f} : A \rightarrow B^{\ast}$ and
$\hat{\omega} : B^{\ast} \rightarrow C$. Their composite,
$\hat{\omega}\circ \hat{f}$, is a positive operator $A \rightarrow
C$. If $ABC$ is a {\em regular} composite, we also have, for any
state $\alpha \in A$ and any effect $c \in C^{\ast}$, that $\alpha
\otimes \omega$ is a state in $ABC$ and $f \otimes c$ is an effect
in $(ABC)^{\ast}$. We now make a technically trivial but crucial
observation:

\begin{lemma} With notation as above, the un-normalized conditional
state of $\alpha \otimes \omega$ given an effect $f \in AB$ is
\[(\alpha^A \otimes \omega^{BC})^{C}_{f} =
\hat{\omega}(\hat{f}(\alpha)).\]
\end{lemma}

\noindent{\em Proof:} As pure tensors generate $BC$, it is
sufficient to check this in the case that $\omega = \beta \otimes
\gamma$. Then, for any $b \in B^{\ast}$, $\hat{\omega}(b) =
\beta(b)\gamma$ (using, here, our convention of identifying a state
space with its double dual). Note also that $f(\alpha \otimes \beta)
= \beta(\hat{f}(\alpha))$. Hence, for any $c \in C^{\ast}$, $(f
\otimes c)(\alpha \otimes \omega) = f(\alpha \otimes \beta)\gamma(c)
= \beta(\hat{f}(\alpha))\gamma(c) =
\widehat{\omega}(\hat{f}(\alpha))(c)$. $\Box$\\

%Expand $\omega$ as a linear combination of pure tensors, say $\omega
%\ = \ \sum_{i} t_i \beta_i \otimes \gamma_i$, where $\beta_i \in
%B_{+}$ and $\gamma_i \in C_+$ for all $i$. Then, for any $\alpha \in
%A$,
%\begin{eqnarray*}
%\ \ (f \otimes Id)(\alpha \otimes \omega) & = & \sum_{i} t_i
%f(\alpha \otimes \beta_i)\gamma_i\\
%& = & \sum_i \hat{f}(\alpha)(\beta_i)\gamma_i\\
%& = & \sum_i t_i \widehat{\beta_i \otimes
%\gamma_i}(\hat{f}(\alpha))\\
%& = & \hat{\omega}(\hat{f}(\alpha))  \ \Box \end{eqnarray*}

\begin{corollary} For any state spaces $A$, $B$ and $C$,
\begin{itemize}
\item[(i)] There is a canonical embedding \[A \mintensor (B \maxtensor C)
\leq (A \mintensor B) \maxtensor C.\] \item[(ii)] The composite $(A
\mintensor B) \maxtensor C  $ is regular.\end{itemize}
\end{corollary}

\noindent {\em Proof:} By Lemma 2, any product state $\alpha \otimes
\omega$ with $\alpha \in A_{+}$ and $\omega \in (B \maxtensor
C)_{+}$ yields a positive bilinear form on $(A \mintensor B)^{\ast}
\times C^{\ast}$, namely, $(\alpha \otimes \beta)(f,c) =
c(\hat{\omega}(\hat{f}(\alpha)))$. Hence, we have a natural positive
linear mapping $A \mintensor (B \maxtensor C) \rightarrow ((A
\mintensor B)^{\ast} \mintensor C^{\ast})^{\ast}$; the last is
isomorphic to $(A \mintensor B) \maxtensor C$. This establishes (i).

To show that $(A \mintensor B) \maxtensor C$ is regular, we first
observe that $BC = B \maxtensor C$. Indeed, let $\mu \in B
\maxtensor C$, and let $\hat{\mu}$ be the associated positive
operator $B^{\ast} \rightarrow C$. Let $\alpha$ be some fixed state
in $A$. Given $f \in (A \mintensor B)^{\ast} \simeq {\cal L}_{+}(A,
B^{\ast})$ and $c \in C^{\ast}$, set
 \[\omega(f,c) =
 \hat{\mu}(\hat{f}(\alpha))(c):\] this is bilinear in $f$ and in
$c$, and positive where both $f$ and $c$ are positive, and so,
defines an element $\omega \in (A \mintensor B) \maxtensor C$. We
now observe that the reduced state $\omega^{BC}_{u_{A}}$, evaluated
on a pair of effects $(b,c) \in B^{\ast} \times C^{\ast}$, yields
\begin{eqnarray*}
\omega^{BC}_{u_{A}}(b,c) & = & \omega(u_A,b,c)\\
& = & \hat{\mu}(\widehat{(u_{A} \otimes b)}(\alpha))(c)\\
&  = &  \hat{\mu}(b)(c) = \mu(b,c).\end{eqnarray*} Thus,
$\omega^{BC}_{u_A} = \mu$. This shows that $B \maxtensor C \leq BC$;
the reverse inclusion is trivial, so $BC \simeq B \maxtensor C$, as
claimed. We now have, by part (i), that \begin{eqnarray*} A
\mintensor (BC) & = & A \mintensor (B \maxtensor C) \\
& \leq & (A \mintensor B) \maxtensor C = ABC.\end{eqnarray*}
Obviously, we have $ABC \leq A \maxtensor (B \maxtensor C) = A
\maxtensor BC$. The corresponding result for the coarse-graining
$\{\{AC\}, \{B\}\}$ follows similarly (or by symmetry), and that for
$\{\{A\},\{B,C\}\}$ is trivial, so so $ABC$ is regular. $\Box$\\

We can interpret Lemma 2 in information-processing terms as follows.
Suppose two parties, Alice and Bob, have access to systems $A$ and
$B$, respectively. Suppose, moreover, that Alice's system consists
of two subsystems, $A_1$ and $A_2$, with $A_1$ in an unknown state
$\alpha$. If the total Alice-Bob system is represented by a regular
composite $AB = A_1A_2B$, then if $f$ is an effect on $A$ and
$\omega$ is a known state on $A_2B$, we may prepare $A_1 A_2 B$ in
the joint state $\alpha \otimes \omega$: if Alice performs a
measurement on $A = A_1 A_2$ having $f$ as a possible outcome, then,
conditional upon securing this outcome, the conditional state of $B$
is, up to normalization, $\hat{\omega}(\hat{f}(\alpha))$. Thus, we
may say that Alice has evaluated a known mapping, namely
$\hat{\omega} \circ \hat{f}$, on an unknown input $\alpha$, simply
by securing $f$ as a measurement outcome. In the sequel, we refer to
this protocol as {\em remote evaluation.}

This is obviously reminiscent of a teleportation protocol. Indeed,
conclusive teleportation can be regarded as the special case of
remote evaluation in which the mapping $\hat{\omega} \circ \hat{f}$
is invertible. Suppose that $\eta: A_1 \simeq B$ is a fixed
isomorphism between Alice's system $A_1$ and Bob's system $B$
(allowing us to say what we mean by saying a state of $B$ is the
same as a state of $A_1$). Suppose, further, that the unknown state
$\alpha$ is recoverable from the {\em normalized} conditional state
$\norm{(\alpha \otimes \omega)}^{B}_{f}$ by means of a physically
admissible process $\tau$, depending on $f$ but not on $\alpha$:
upon securing a measurement outcome corresponding to $f$, Alice can
then instruct Bob to make the correction $\tau$; once this is done,
she is certain that the conditional state of Bob's system $B$
--- whatever it is --- is identical (up to $\eta$) to the original, but unknown,
state $\alpha$.

In fact, we can distinguish two situations: one in which the
correction operation $\tau$ is certain to succeed, and another in
which it may fail, but in which this failure will be apparent to
Bob. In the latter case, the teleportation protocol has an
additional step: Alice must wait for Bob to report the success of
the correction. We shall refer to these as {\em strong}
and {\em weak} conclusive teleportation, respectively. Notice that
the standard (one-outcome post-selected) quantum teleportation
protocol is an instance of a strong teleportation protocol.

We make this language precise as follows. To avoid needess
repetition, here and in the balance of this paper $A_1 A_2 B$
denotes a regular composite of state spaces $A_1, A_2$ and $B$ with
$A_1$ isomorphic to $B$ by a fixed isomorphism $\eta : A_1 \simeq
B$; and $f$ is an effect on $A_1 A_2$ and $\omega$ is a state in
$A_2 B$.

\begin{definition}
We say that the pair $(f,\omega)$ is a {\em conclusive teleportation
protocol} on $A_1A_2B$ iff there exists a norm-contractive linear
mapping $\tau : B \rightarrow B$, called a {\em correction}, such
that, for every normalized state $\alpha \in \Omega_{A_1}$,
 \[\tau(\norm{(\alpha \otimes \omega)}^{B}_{f}) =
 t_{\alpha} \eta(\alpha)\]
for some constant $t_{\alpha} > 0$.  If $\tau$ can be so chosen that
$t_{\alpha} = 1$ for all $\alpha$, we say that the protocol $(f, \omega)$ is {\em
strong}.
\end{definition}

By Lemma 2, the conditional state $\widetilde{(\alpha \otimes
\omega)}^{B}_{f}$ can be expressed as
$\hat{\omega}(\hat{f}(\alpha))/u(\hat{\omega}(\hat{f}(\alpha)))$.
Let
\[\mu  := \hat{\omega} \circ \hat{f} : A_1 \rightarrow B,\]
noting that this is a norm-contractive positive mapping. Then
$(f,\omega)$ is a teleportation protocol iff there exists a
norm-contractive positive mapping $\tau : B \rightarrow B$ with
\[\tau(\mu(\alpha)) = t_{\alpha}\|\mu(\alpha)\|\eta(\alpha)\]
for all $\alpha \in \Omega_{A}$.
Notice that $t_{\alpha} = u(\tau(\norm{\mu(\alpha)}))$, i.e.,
$t_{\alpha}$ is the probability that the correction $\tau$ {\em
succeeds} in the conditional state $\norm{\mu(\alpha)})$.
Accordingly, a strong protocol is one for which there exists a
correction that is certain to succeed.

Theorem 1, below, gives a complete characterization of conclusive
teleportation protocols, strong or otherwise, in terms of the
mapping $\mu = \hat{\omega} \circ \hat{f}$. We require an easy
preliminary

\begin{lemma} Let $A$ and $B$ be any abstract state spaces. Let
$\phi, \psi : A \rightarrow B$ be any two linear mappings with
$\psi$ injective.  If, for every $\alpha \in \Omega_{A}$, there is a
constant $k(\alpha)$ such that $\phi(\alpha) =
k(\alpha)\psi(\alpha)$, then in fact $k(\alpha) \equiv k$, a
constant not depending on $\alpha$.
\end{lemma}

\noindent{\em Proof:}  Let $\alpha$ and $\beta$ be distinct, and
hence, linearly independent, elements of $\Omega_{A}$, and consider
$\gamma = (\alpha + \beta)/2$. Then we have
\[\phi(\gamma) = k(\gamma)\psi(\gamma) = (k(\gamma)/2)(\phi(\alpha) +
\phi(\beta))\] and also
\[\phi(\gamma) = (\phi(\alpha) + \phi(\beta))/2 =
(k(\alpha)\psi(\alpha) + k(\beta)\psi(\beta) )/2.\] Thus,
\[(k(\alpha) - k(\gamma))\psi(\alpha) + (k(\beta) -
k(\gamma))\psi(\beta) = 0.\] Since $\psi$ is injective,
$\psi(\alpha)$ and $\psi(\beta)$ are linearly independent in $B$;
hence, $k(\alpha) - k(\gamma) = k(\beta) - k(\gamma) = 0$, whence
$k(\alpha) = k(\beta)$. $\Box$\\

Recall that an {\em order-isomorphism} between abstract state spaces
is a positive linear bijection with a positive inverse, while an
{\em isomorphism} also preserves normalization.

\begin{theorem} Let $\mu := \hat{\omega} \circ \hat{f} : A_1
\rightarrow B$. Then
\begin{itemize}
\item[(a)] $(f,\omega)$ is a conclusive teleportation protocol iff
$\mu$ is an order isomorphism; in this case $\tau = s(\eta \circ
\mu^{-1})$ where $s \leq 1/\|\mu^{-1}\| \leq 1$.
\item[(b)] $(f,\omega)$  is a strong teleportation protocol iff $\mu$ is proportional to an isomorphism; in this
case, the correction $\tau$ is a symmetry of $B$.
\end{itemize}
\end{theorem}

\noindent{\em Proof:}

(a) Suppose first that $(f,\omega)$ is a teleportation protocol.
Then there exists a positive, norm-contractive mapping $\tau : B
\rightarrow B$ such that, for all $\alpha \in \Omega_{A_{1}}$, there
\[\tau(\mu(\alpha)) = t_{\alpha} \|\mu(\alpha)\| \eta(\alpha)\]
for some constant $t_{\alpha} > 0$. As $\eta$ is injective, Lemma 4
implies that
\[t_{\alpha} \|\mu(\alpha)\| \equiv s,\]
a constant independent of $\alpha$. Note that, as $\tau$ is
norm-contractive, $s < 1$. Since $\Omega_{A_1}$ spans $A_1$, we have
$\tau \circ \mu = s \eta$. It follows that $\tau : B \rightarrow B$
is a surjective linear mapping. As we are working in finite
dimensions, this implies that $\tau$ is invertible; we have
\[\tau^{-1} = \mu \circ \frac{1}{s} \eta^{-1},\]
which is positive. Thus, $\tau$ is an order-isomorphism. It follows
$\mu = \tau^{-1} \circ s \eta$ is also an order-isomorphism.

For the converse, suppose that $\mu$ is an order-isomorphism. Then
$\eta \circ \mu^{-1}$ is also an order-isomorphism. Let
\[\tau := s (\eta \circ \mu^{-1})\]
where $s < 1/\|\mu^{-1}\|$. As $\|\eta\| = 1$, we have $\|\tau\|
\leq s \|\eta\|\|\mu^{-1}\| < 1$, so $\tau$ is norm-contractive. Now
$\tau \circ \mu  = s \eta$. For all $\alpha$, let $t_{\alpha} =
s/\|\mu(\alpha)\|$ (noting that $\|\mu(\alpha)\| > 0$, since $\mu$
is injective), so that
\[\tau(\mu(\alpha)) = s\eta(\alpha) = t_{\alpha} \|\mu(\alpha)\|
\eta(\alpha).\]

(b) Suppose first that $\mu = k\phi$ for some isomorphism $\phi :
A_1 \rightarrow B$. Then $\phi : A_1 \rightarrow B$ and some
positive constant $k$. Let $\tau = \eta \circ \phi^{-1}$: then
$\tau(\mu(\alpha)) = k\eta(\alpha)$ for all $\alpha$. Since $k =
\|\mu(\alpha)\|$ for all $\alpha$, we have a strong teleportation
protocol.

For the converse, suppose $(f,\omega)$ is a strong teleportation
protocol. Thus, there exists a norm-contractive positive mapping
$\tau : B \rightarrow B$ such that, for all $\alpha \in
\Omega_{A_1}$,
\[\tau(\mu(\alpha)) = \|\mu(\alpha)\|\eta(\alpha).\]
We claim that $\tau$ is a symmetry. To see this, let $\Gamma =
\mu(\Omega_{A}) := \{ \mu(\alpha) | \alpha \in \Omega_{A_1}\}$, and
set $\norm{\Gamma} = \{\norm{\gamma} | \gamma \in \Gamma\}$; note
that this set is a convex subset of $\Omega_{B}$.\footnote{To spell
this out, suppose $\gamma_1, \gamma_2 \in \Gamma$ with normalized
versions $\tilde{\gamma}_{1} = t_1 \gamma_1, \tilde{\gamma}_{2} =
t_2 \gamma_{2} \in \tilde{\Gamma}$. Now consider a convex
combination
\[\alpha = p\norm{\gamma}_1 + q \norm{\gamma}_2\]
where $p, q \geq 0$ with $p + q = 1$. Then
\[\alpha = pt_1 \gamma_1 + q t_2 \gamma_2.\] Let
\[\gamma = \frac{pt_1}{pt_1 + q t_2} \gamma_1 + \frac{pt_2}{pt_1 +
qt_2} \gamma_2 \in \Gamma\] and note that
\[\alpha = (pt_1 + q t_2) \gamma = \norm{\gamma} \in \norm{\Gamma}.\]}
Now, $\tau(\tilde{\mu}(\alpha)) = \eta(\alpha) \in \Omega_{B}$, so
$\tau$ effects an affine bijection of $\Gamma$ onto $\Omega_{B}$. It
follows that the affine span of $\Gamma$ equals that of $\Omega$,
whence, that $\tau$ preserves the affine span of the latter -- which
is exactly the hyperplane $u_{B}^{-1}(1)$. As $\tau$ is positive, it
also preserves the positive cone $B_{+}$, whence, $\tau$ preserves
$B_{+} \cap u^{-1}(1) = \Omega_{B}$. Thus, $\tau$ is a symmetry, as
claimed.   It remains to show that $\mu$ is proportional to an
isomorphism. But as we have
\[\tau(\mu(\alpha)) = \|\mu(\alpha)\|
\eta(\alpha),\] we also have
\[\mu(\alpha) = \|\mu(\alpha)\| \tau^{-1}(\eta(\alpha))\]
for all $\alpha \in \Omega_{A_1}$. Invoking Lemma 4, we see that
$\|\mu(\alpha)\| \equiv k$, a constant independent of $\alpha$ --
whence, $\mu = k \tau^{-1}\circ \eta$. $\Box$\\

\noindent{\em Remarks:} (1) For Bob to be able to apply the
correction mapping $\tau$, the latter must belong to the dynamical
semigroup ${\frak D}_{B}$. Given our simplifying assumption is that
${\frak D}_{B}$ comprises all norm-contractive positive mappings on
$B$, this is automatic, but in a treatment using more general
dynamical models, it would need to be assumed as part of the
definition of a teleportation protocol.

(2) If $(f,\omega)$ is a teleportation protocol on $A_1 A_2 B$, then
we can regard it also as a teleportation protocol on $A_1 \mintensor
(A_2 \maxtensor B)$, as the latter is regular, $f$ is an effect on
$A_1 \mintensor A_2$, and $\omega$ is a state in $A_2 \maxtensor B$.
Thus, all teleportation protocols involving regular
composites of $A_1, A_2$ and $B$ live, so to speak, in
$A_2 \mintensor (A_2 \maxtensor B)$. \\

One can regard non-strong conclusive teleportation protocols as {\em
inherently inefficient}. The question arises, whether an inefficient
protocol can always be replaced with one that is perfectly
efficient. We show that this is always possible when the composite
is dynamically admissible. (Recall under our standing assumption,
a composite is
dynamically admissible iff its positive cone is closed under
products of positive mappings on the factors.)

\begin{corollary} Suppose $A_1 A_2 B$ is dynamically
admissible. If $(f,\omega)$ is a conclusive teleportation protocol
with correction $\tau$, then let $\omega' \in A_2B$ be the state
defined, for all $a \in A_2^{\ast}$ and $b \in B^{\ast}$, by
\[\omega'(a,b) = \norm{\omega(a, \tau(b))}.\]
Then $(f,\omega')$ is a strong conclusive teleportation protocol,
requiring no correction. \end{corollary}

\noindent{\em Proof:} Since $(f,\omega)$ is a conclusive
teleportation protocol, there exists a positive mapping $\tau : B
\rightarrow B$ such that \[\tau \circ \hat{\omega} \circ \hat{f} = s
\eta\] for some constant $s$. Since $A_1 A_2 B$ is dynamically
admissible, $\omega' \in A_2B$. It is easily verified that
$\hat{\omega}' = (\tau \circ \omega)/\|\tau \circ \omega\|$; hence,
\[  \hat{\omega}' \circ \hat{f} = \frac{s}{\|\tau \circ \omega\|}
\eta.\] Thus, $\hat{\mu}' := \hat{\omega}' \circ \hat{f}$ is
proportional to a symmetry, so $(f,\hat{\omega})$ is a strong
conclusive teleportation protocol, by Theorem 1. Moreover, as the
symmetry in question is $\eta$ itself, no correction is required.
$\Box$ \\

%(2)  There is a temptation to compose $\hat{\omega}$ with the
%correction, $\tau$, to obtain a new state $\omega' := \tau \circ
%\hat{\omega}$; the resulting teleportation protocol $(f,\omega')$
%would then be strong (by Theorem 1), and require no correction.
%However, this assumes that the normalized tripartite state space
%$\Omega_{A_1A_2B}$ is invariant under the local action of $B$'s
%automorphism group on the third factor. This is a nontrivial
%assumption, which we prefer not to pursue here.

%\begin{corollary} In order that a regular composite of $A_1, A_2$
%and $B \simeq A_1$ support a conclusive (resp., strong conclusive)
%teleportation protocol, it is necessary and sufficient that there
%exist positive elements $\omega \in A_{2}B$ and $f \in A_1A_2$ such
%that $\hat{\omega} \circ \hat{f} : A_1 \rightarrow B$ is an
%order-isomorphism (resp., an isomorphism).
%\end{corollary}

%\noindent{\em Proof:} Necessity is obvious. For sufficiency, suppose
%$f$ and $\omega$ are as indicated. Then there exist constants $n, m
%> 0$ such that $n\omega$ and $mf$ are respectively a state and an
%effect. We then have \[nm = nm\|\hat{\omega} \circ \hat{f}\| =
%\|n\omega \circ m f \| \leq 1,\] so setting $k = nm$ gives the
%desired result. $\Box$       [Check!]                               \\

It follows from Theorem 1 that if a bipartite state $\omega$ on $A_2
B$ and a bipartite state $f$ on $A_1 A_2$ supply a conclusive
teleportation protocol, then the positive linear mappings $\hat{f}$
and $\hat{\omega}$ are respectively injective and surjective. We can
be somewhat more precise about the geometry of the situation. Let us
say that a {\em compression} on an ordered space $V$ is a positive
mapping $P : V \rightarrow V$ such that $P^2 = P$. Equivalently,
$P$'s range, $P(V)$, is an ordered subspace of $V$, and $P(\alpha) =
\alpha$ for all $\alpha \in P(V)$. As an example, let $K$ be a cube,
and let $F$ be a face thereof; the obvious affine surjection $K
\rightarrow F$ extends to a compression $V(K) \rightarrow V(F)$.

Suppose now that $(f,\omega)$ is a conclusive teleportation protocol
on $A_1 A_2 B$ with an order-isomorphic correction $\tau : B
\rightarrow B$, so that
\[ \tau \circ \hat{\omega} \circ \hat{f} = s \eta\]
for some constant $s > 0$. Then $\hat{f} : A_1 \rightarrow
A_{2}^{\ast}$ is an order-embedding, and and $\hat{\omega} :
A_{2}^{\ast} \rightarrow B$ is a positive surjection. Let \[P : =
\hat{f} \circ \eta^{-1} \circ \tau \circ \hat{\omega} : A_{2}^{\ast}
\rightarrow A_{2}^{\ast} :\] an easy computation shows that $P$ is a
compression in the above-defined sense, with range equal to the
image of $\hat{f}$.

Conversely, suppose we are given an effect $f$ such that
$\hat{f} : A_{1} \rightarrow A_{2}^{\ast}$ taking $A_{1}$
order-isomorphically onto the range of a compression $P :
A_{2}^{\ast} \rightarrow A_{2}^{\ast}$. Let $\hat{f}^{+} : \ran(P)
\rightarrow A_1$ be the inverse of $\hat{f}$'s co-restriction to
$\ran(P)$, and let $\alpha_o = \hat{f}^{+}(u_{A_2})$, i.e, the
unique element of $A_{1 +}$ such that $\hat{f}(\alpha_o) =
P(u_{A_2})$. Define
\[\hat{\omega'} := \frac{1}{\|\alpha_o\|} \eta \circ \hat{f}^{+} \circ
P.\] Then $\hat{\omega'}(u_{A_2}) = \eta(\alpha)/\|\alpha_o\| \in
\Omega_{B}$ (since $\eta$ is an isomorphism, hence norm-preserving),
whence, $\hat{\omega'}$ corresponds to a normalized state $\omega'$
in $A_2 \maxtensor B$. The pair $(f,\omega')$ gives us a {\em
strong}---and correction-free---teleportation protocol on $A_1
\mintensor (A_2 \maxtensor B)$. If $A_1 A_2 B$ is dynamically
admissible, then $\omega' \in A_2 B$, and indeed, is precisely the
state $\omega'$ defined in Corollary 2.

Summarizing:
%as a sub-ordered space of $A_{2}^{\ast}$. and isomorphisms $\hat{f}$
%and $\eta$ as above, with $f \in (AB)^{\ast}$ and $\eta \circ P =
%\hat{\omega}$ for some $\omega \in A_2B$, the pair $(f, \omega)$
%constitutes a conclusive teleportation protocol. Similar statements
%hold for strong conclusive teleportation, requiring that [FIX!]
%Summarizing:

\begin{theorem} Let $A_1, A_2$ and $B \simeq A_1$ be abstract
state spaces with $B \simeq A_1$.
A regular composite $A_1A_2B$ supports a conclusive teleportation
protocol iff there exists an effect $f$ on $A_1 A_2$, a state $\omega$
in $A_2 B$, and a compression $P : A_2^{\ast} \rightarrow
A_{2}^{\ast}$ such that $\hat{f}$, co-restricted to $\ran(P)$, is an
order-isomorphism $A_1 \simeq \ran(P)$ and $\hat{\omega}$, restricted
to $\ran(P)$, is an order-isomorphism $\ran(P) \simeq B$.
\end{theorem}
%Maybe lemma should be: have CT iff have state and effect with
%$\hat{f}\circ \hat{\omega} = r\Id$ for some $r > 0$.

%Theorem 1 supplies a wealth of examples of tripartite systems
%supporting conclusive teleportation. Let us say that a {\em
%compression} on an ordered space $V$ is a positive mapping $P : V
%\rightarrow V$ such that $P^2 = P$. Equivalently, $P$'s range,
%$P(V)$, is an ordered subspace of $V$, and $P(\alpha) = \alpha$ for
%all $\alpha \in P(V)$. As an example, let $K$ be a cube, and let $F$
%be a face thereof; the obvious affine surjection $K \rightarrow F$
%extends to a compression $V(K) \rightarrow V(F)$.

\begin{corollary} $A_1 \mintensor (A_2 \maxtensor A_1)$ supports conclusive
teleportation with $\eta(\alpha) = \alpha$ for all $\alpha$ iff $A_1
\leq A_{2}^{\ast}$ is the range of a compression $P : A_{2}^{\ast}
\rightarrow A_{2}^{\ast}$.
\end{corollary}

\noindent{\em Proof:} Suppose first that we have a compression $P:
A_{2}^{\ast} \rightarrow A_{1}^{\ast}$: regarding $P$ as a positive
surjection $\pi : A_{2}^{\ast} \rightarrow A_1$, and letting $\iota
: A_1 \rightarrow A_{2}^{\ast}$ be the positive inclusion mapping,
we have $\pi \in \widehat{(A_2 \maxtensor A_1)}$ and $\iota \in
\widehat{(A_1 \mintensor A_2)^{\ast}}$. As $(\pi \circ \iota)
(\alpha) = \alpha$ for all $\alpha \in A_1$, Theorem 1 tells us that
$A_1 A_2 A_1  = A_1 \mintensor (A_2 \maxtensor A_1)$ supports
conclusive teleportation.

Conversely, if $A_1 \mintensor (A_2 \maxtensor A_1)$ supports
conclusive teleportation, then by Corollary 2, there exist positive
operators $\hat{\omega}: A_{2}^{\ast} \rightarrow A_{1}$ and
$\hat{f}: A_1 \rightarrow A_{2}^{\ast}$ with $\hat{\omega} \circ
\hat{f} : A_1 \rightarrow A_1$ an isomorphism, in which case $P :=
\hat{f} \circ \hat{\omega}$ is a compression. $\Box$  \\

%%What about sub-normalization??

We also have

\begin{corollary} Let $A_1 A_2 B$ be a regular composite of three pairwise
isomorphic, weakly self-dual state spaces. If $A_2 B$ contains a
state $\omega$ with $\hat{\omega} : A_2^{\ast} \simeq B$, then $A_1
A_2 B$ supports conclusive teleportation. In particular, $A_1
\mintensor (A_2 \maxtensor B)$ supports conclusive teleportation.
\end{corollary}

{\em Remark:} As observed above, the standing assumption that for a
system $A$, its dynamical semigroup $\fD_A$ is the set of all
positive maps on $A$, strongly restricts the nature of dynamically
admissible tensor products, and is, for example, incompatible with
the usual quantum tensor product. However, our definitions and
results concerning teleportation are easily adapted to the setting
of regular composites of arbitrary dynamical models: as noted above,
the definition of a teleportation protocol in that setting requires
that the correction mapping $\tau^{B}$ on $B$  belong to the
dynamical semigroup ${\frak D}_{B}$; with this modification, one has
one has obvious analogues of Theorems 1 and 2, and of Corollary 2.
\\

\noindent{\bf 4. Deterministic Teleportation} As in the previous
section, $A_1A_2 B$ is a regular composite of three state spaces
$A_1$, $A_2$ and $B$, with $B$ isomorphic to $A_1$. In order for
$A_1 A_2 B$ to support a {\em deterministic} teleportation protocol,
we require a bipartite state $\omega \in A_2B$ and an {\em
observable} $\{f_1,...,f_n\}$ on $A_1 A_2$ such that for every state
$\alpha$ in $A_{1}$ and for each $i$, the state $\alpha$ is
recoverable from the conditional state of $\alpha \otimes \omega$
given outcome (effect) $f_i$.

\begin{definition}
Let $A_1 A_2 B$ be a regular composite of $A_1$, $A_2$ and $B$ with
$B \simeq A_1$ via a fixed isomorphism $\eta : A_1 \rightarrow B$.
If $\omega$ is a state in $A_2 B$ and $E = (f_1,...,f_n)$ is an
observable on $A = A_1 A_2$, we shall say that the pair $(E,
\omega)$ realizes a {\em deterministic teleportation protocol} iff,
for each effect $f_i \in E$, the pair $(f_i, \omega)$ realizes a
%deterministic \
strong conclusive teleportation protocol.
\end{definition}

The idea is that, upon measuring $E$ and obtaining outcome $f_i$,
Alice instructs Bob to apply a suitable correction $\tau_i$; the
conditional state of $B$ is then $\eta(\alpha)$. Note that, by
Theorem 1, the correction $\tau_i$ must be a symmetry of $B$.

At present, it is not clear to us exactly what conditions on the
pair $A_1, A_2$ will be necessary in order to secure a deterministic
teleportation protocol. However, Theorem 2 below provides a wealth
of examples of systems which, while weakly self-dual, are neither
classical nor quantum, but can nevertheless by combined so as to
support a deterministic teleportation protocol. In particular,
self-duality is not necessary for deterministic teleportation.

In what follows, let $A$ be an abstract state space carrying an
action of a finite group $G$ that preserves the state space
$\Omega$. Note that there is a canonical dual action of $G$ on
$A^{\ast}$ given by
\[(ga)(\alpha) = a(g^{-1} \alpha)\]
for all $g \in G$, $a \in A^{\ast}$, and  $\alpha \in A$. Note, too,
that the order-unit $u = u_{A}$ is invariant under this action,
i.e,. $gu = u$ for all $g \in G$.   A state $\omega$ is called
$G$-{\em equivariant} if for all $g \in G$ and all effects $a \in
A^{\ast}$ we have \beq g \hat{\omega}(a) = \hat{\omega}(ga)\;. \eeq
%%Awkward that we are using $A$ here for $A_1$, where $A$ formerly stood for $A_1 A_2$.

\begin{theorem} Let $A$ be weakly self-dual, and suppose $G$ is a finite group acting on
$A$, in such a way that (i) $G$ acts transitively on the extreme
points of $\Omega$, and (ii) there exists a $G$-equivariant
isomorphism $A^{\ast} \simeq A$. Then $A \mintensor (A \maxtensor
A)$ supports a deterministic teleportation protocol. \end{theorem}

For an example, consider the state space obtained by taking $\Omega$
to be a unit square in ${\Bbb R}^{3}$, displaced one unit from the
origin; $A_{+}$ is the cone generated by this square base. As
observed earlier, with respect to the usual inner product,
$A^{\ast}$ can be represented as ${\Bbb R}^{3}$ with cone obtained
by rotating $A_{+}$ by $\pi/4$. This gives us an order-isomorphism
$A^{\ast} \rightarrow A$ that is equivariant with respect to the the
natural action of ${\Bbb Z}_{4}$ on $\Omega$; as this last is
transitive on the vertices of the latter, Theorem 2 tells us that $A
\mintensor (A \maxtensor A)$ will support a deterministic
teleportation protocol. Similar considerations show that the same
conclusion holds whenever $\Omega_A$ is any regular polygon.             \\

For the proof of Theorem 3, we need an easy lemma.

\begin{lemma} Let $A$ and $G$ be as in Theorem 3.
Then there exists a unique invariant normalized state $\omega_o \in
\Omega_{A}$. \end{lemma}

\noindent{\em Proof:} Notice, first, that there is certainly at
least one fixed state, namely
$(1/|G|) \omega_o = \sum_{g \in G} g\alpha_o$, where $\alpha_o$ is any one extreme state.
%and $\mu$ is the normalized Haar measure on $G$.
To see that there can be no more
than one such state, let $\Gamma$ denote the set of $G$-fixed points
of $\Omega$. Observe that $\Gamma$ is an affine section of $\Omega$;
hence, if $\Gamma$ contains more than a single point, it contains an
affine line, which must intersect the topological boundary of
$\Omega$. Let $\alpha$ be a fixed state belonging to this boundary:
equivalently, $\alpha$ is fixed, and belongs to a proper face of
$\Omega$. Let $F$ be the smallest face containing $\alpha$: for each
$g \in G$, $gF$ is again a face containing $\alpha$, so $F \subseteq
gF$. In other words, $F$ is invariant. But since $F$ is a proper
face and $G$ acts transitively on $\Omega$'s extreme points, this is
impossible. $\Box$\\

\noindent{\em Proof of Theorem 3:} Let $A$, $G$ and $\omega_o$ be as
above. By assumption, there is an equivariant order-isomorphism
$\phi : A^{\ast} \rightarrow A$; normalizing if necessary, we can
assume that $\phi = \hat{\omega}$ for some bipartite state on $AB$.
We claim that $\hat{\omega}(u) = \omega_o$. Indeed, for all $g \in
G$, %%and for all effects $e \in A^{\ast}$,
we have
\[g\hat{\omega}(u) = \hat{\omega}(gu) = \hat{\omega}(u).\] Thus,
$\hat{\omega}(u)$ is $G$-invariant; but there is only one invariant
state, namely $\omega_o$.

Now, for all $g \in G$, let $f_{g} \in (A \maxtensor A)^{\ast}$
correspond to the operator
\[\hat{f}_{g} = \frac{1}{|G|} \hat{\omega}^{-1} \circ g.\]
We claim that  $E = \{f_{g}\}$ is an observable, and $(E, \omega_o)$
realizes a strong deterministic teleportation protocol. To see this,
note that for every $\alpha \in A$, $\frac{1}{|G|}\sum_{g \in G}
g\alpha$ is a $G$-invariant state, and hence, by Lemma 3,
equals $\omega_o$. Thus,
\begin{eqnarray*} \sum_{g \in G} f_{g}(\alpha)
& = & \sum_{g \in G} \frac{1}{|G|} \hat{\omega}^{-1}( g\alpha)\\
& = & \hat{\omega}^{-1}\left (\frac{1}{|G|} \sum_{g \in G} g\alpha \right )\\
& = & \hat{\omega}^{-1}(\omega_o) = u\end{eqnarray*} (appealing, in
the last step, to the fact that $\hat{\omega}(u) = \omega_o$). So
$\sum_{g \in G} f_{g} = u$, i.e., $g \mapsto f_{g}$ is an
observable. Moreover,
\[\hat{\omega}(\hat{f_g}(\alpha)) =
\hat{\omega}(\hat{\omega}^{-1}(g\alpha)) = g\alpha.\] Thus,
$\hat{\omega} \circ \hat{f}_g$ acts as the group element $g \in G$
--  and hence, in particular, has a norm-preserving inverse.
$\Box$\\

%Theorem 2 yields a wealth of examples %%Phrasing repetitious of above... Put example above!
%of weakly self-dual, but not self-dual, state spaces supporting
%deterministic teleportation
%protocols. %(at least if we construct our tripartite systems using the mixed min/max tensor product construction discussed above.)
%Indeed, the case in which $\Omega$
%is a regular polygon %[check: or any regular polytope?]
%would furnish such an example. \\%%Discuss example of a square-based cone here...\\

%\noindent{\bf Example:} Consider the system whose state space
%$\Omega$ is a square. Concretely, we can take this to be the state
%space of a test space ${\frak A} = \{\{a,a'\}, \{b,b'\}\}$
%consisting of two disjoint, two-outcome tests). Let $G$ be the
%cyclic group ${\Bbb Z}_4$, acting by rotation. The unique
%$G$-invariant state is the center of the square (the uniform state
%assigning probability 1/2 to every outcome). The dual cone is
%generated by four co-vectors corresponding to the four outcomes of
%${\frak A}$. In an appropriate representation, these form the
%vertices of another square, with orthogonal outcomes corresponding
%to diagonally opposite vertices. This allows us easily to construct
%an equivariant order-isomorphism $A^{\ast} \rightarrow A$.\\

\noindent{\em Remarks:} If the group $G$ is compact, we can replace
the discrete observable $\{f_{g} | g \in G\}$ in Theorem 2 by the
continuous $G$-valued density $g \mapsto f_{g} := \int_{G}
\omega^{-1} ~ \circ g ~d\mu(g)$, where $\mu$ is the normalized Haar
measure on $G$. While it is far from clear that we should want to
regard this as a ``continuously indexed observable" in any literal
sense, it may be that discrete, coarse-grained versions of the
effect-valued measure $B \mapsto \int_{g \in B} f_{g} d\mu(g)$ ($B$
ranging over Borel subsets of $G$) can each underwrite some form of
approximate teleportation protocol, of which a deterministic
protocol is in some sense the limiting case. We defer exploration of this possibility
to a future paper.

Also note that homogeneity of $A$ implies that the group of base-preserving
automorphisms of $A$, which is finite or compact, acts transitively on the extreme
points of $\Omega_A$, so homogeneous weakly self-dual state spaces are good
candidates for supporting the deterministic teleportation protocol described
in Theorem 2, or its continuous analogue. \\

%(2) To get a really sharp theorem, we need some reasonable
%conditions that guarantee the existence of a bipartite state
%$\omega$ with $\hat{\omega}$ an equivariant isomorphism. Here's a
%stab at this: Let $\phi : A \rightarrow A^{\ast}$ be any fixed
%order-isomorphism. Let $G \times G$ act on ${\cal L}(A)$ by
%$(g,h)\phi = g \phi h^{-1}$. There exists an open ball $B$ around
%$\phi$ consisting of order-isomorphisms. The orbit of this ball
%under $G \times G$ consists of finitely many open balls containing
%$\phi$; the intersection of these yields an open, convex,
%$G$-invariant subset $C$ of ${\cal L}(A)$, consisting entirely of
%order-isomorphisms. The set of all strictly positive multiples of
%elements of $C$ is again convex, open, $G$-invariant, and consists
%of order-isomorphsims. The closure of this set is (I believe) an
%invariant cone, the interior of which is the open cone of
%order-automorphisms we began with (use the continuity of the
%determinant for this).  By Corollary 3.5 of Horn (Lin. Alg. Appl.
%1978), there exists a fixed point of $G$ in the interior of this
%cone. This will be the desired equivariant state.

%{\em Question:} Can we define an associative tensor product for
%a class of weakly self-dual state spaces that preserves the property
%that the pure states are permuted transitively by the group of
%order-automorphisms?

\noindent{\bf 5. Entanglement Swapping } Consider a scenario in
which Alice and Bob each possess one wing of two non-local,
bipartite systems, say $S_1 = A_1B_1$ and $S_2 = A_2B_2$. We may
model this situation by supposing that the total system, $S$, is a
composite of the four components $A_1, A_2, B_1$ and $B_2$. We then
have, in addition to the two non-local marginal systems $S_1$ and
$S_2$, two local systems,  $A = A_1 A_2$ and $B = B_1 B_2$
corresponding to Alice and Bob, respectively.

Suppose now that $f$ is an effect on $A = A_1 A_2$ and $\mu$ and
$\omega$ are states in $S_1 = A_1B_1$ and $S_2 = A_2B_2$,
respectively. We have corresponding positive operators $\hat{f} :
A_{1} \rightarrow A_{2}^{\ast}$, $\hat{\omega} : A_{2}^{\ast}
\rightarrow B_{2}$, and $\hat{\mu}^{\ast} : B_{1}^{\ast} \rightarrow
A_{1}$ (the dual of $\hat{\mu} : A_{1}^{\ast} \rightarrow B_1$).
Composing, we obtain a positive operator $\hat{\omega} \circ \hat{f}
\circ \hat{\mu}^{\ast} : B_{1}^{\ast} \rightarrow B_{2}$,
corresponding to a sub-normalized state in $B_1 \maxtensor B_2$.  The
question arises, does this belong to the marginal state space $B =
B_1B_2$? Equivalently, can we implement the mapping in question by
(un-normalized) conditionalization on the outcome of a measurement
on $A$?

If $S$ is a {\em regular} composite of $A_1, A_2, B_1$ and $B_2$,
the answer is yes: $\mu \otimes \omega$ is then a legitimate state
on $S = AB$, whence,  for all $f \in A^{\ast}$, the partially
evaluated state $(\mu \otimes \omega)_{B}(f) = (\mu \otimes
\omega)(f \otimes \ - \ )$ lies in $B$. Now notice the
following analogue of Lemma 1 (proved in the same way, i.e,. by
checking it on pure tensors):

\begin{lemma} With notation as above,
\[ (f^{A} \otimes g^{B})(\mu^{S_1} \otimes \omega^{S_2}) = g^{B}(\hat{\omega} \circ \hat{f}
\circ \hat{\mu}^{\ast}).\]
\end{lemma}
It follows that
\[\hat{\omega} \circ \hat{f} \circ \hat{\mu}^{\ast} = (\mu \otimes
\omega)_{B}(f) \in B,\] as claimed. This is analogous to the remote
evaluation protocol of Section 3: conditional upon Alice securing a
measurement outcome corresponding to $f^{A}$, the conditional state
of Bob's system $B = B_1 \otimes B_2$  corresponds to the operator
$\hat{\omega} \circ \hat{f^{A}} \circ \hat{\mu}^{\ast}$. We might
call this {\em state-pivoting}, as one can easily verify that the
marginal state of $B_1$ is undisturbed.

Where the operation $\hat{\omega} \circ \hat{f}$ can be reversed,
this protocol can be used to transfer the state $\mu$ from subsystem
$S_1$ to subsystem $B$, as in conventional entanglement-swapping
Indeed, suppose that (i) $A_1 = B_2$, (ii) there exists a conclusive
teleportation protocol for the tripartite system $A_1 A_2 B_2$---
i.e., that we can find a state $\hat{\omega}$ in $S_2$ and an effect
$f$ in $A^{\ast}$  such that $\hat{\omega} \circ \hat{f}$ is
proportional to the identity operator on $A_1$. Then, for any $\mu
\in S_2$, Lemma 4 tells us that
\[(\mu \otimes \omega)_{f} = \mu:\]
That is, conditional on the occurrence of $f$ in some measurement by
Alice on system $A$, the state of Bob's system $B$ is $\mu$. In this
situation, we may say that $\mu$ has been teleported from $S_1$ {\em
through} $\omega$ to $B$.

The same considerations also allow us to convert an effect $f$ on
$A$ into a sub-normalized state on $B$. Indeed, if $S_1$ and $S_2$
contain states $\eta_1$ and $\eta_2$, respectively, corresponding to
order-isomorphisms $\hat{\eta}_i: B_{i}^{\ast} \simeq A_i$ for $i =
1,2$, then the mapping $\hat{f} \mapsto \hat{\eta}_1 \circ \hat{f}
\circ \hat{\eta}^{\ast}_2$ gives us an order-preserving linear
injection from $A^{\ast}$ to $B$. Pursuing this a bit further, let
$({\cal C},\ostar)$ be a monoidal theory, as defined in Section 2.
Let us say that a state-space $A \in {\cal C}$ is {\em ${\cal
C}$-self dual} iff there exists a state $\eta \in A \ostar A$ with
$\hat{\eta} : A^{\ast} \rightarrow A$ an isomorphism, and
$\hat{\eta}^{-1} : A \rightarrow A^{\ast}$ corresponding to an
effect in $(A \ostar A)^{\ast}$. It follows from the above, %awk
with $A_1 = A_2$ and $B_1 = B_2$, that if $A$ and $B$ are ${\cal
C}$-self dual, then so is $A \ostar B$. \\

\noindent{\em Four-part disharmonies} The entanglement-swapping
protocol described above can be applied negatively, to show that
certain four-part composites aren't regular. \\

\noindent{\bf Example:} Consider any four non-classical state spaces
$A_1,A_2,B_1$ and $B_2$ with $B_2 \simeq A_1$. If $A_1, A_2$ and
$B_2$ support a conclusive teleportation protocol (in particular, if
all three are isomorphic and weakly self-dual) then the composite
\[S := (A_1 \mintensor A_2) \maxtensor (B_1 \mintensor B_2)\]
cannot be regular. Indeed, arguing as in the proof of Corollary 1,
we see that the reduced system $B := B_1B_2$ is precisely $B_1
\mintensor B_2 \simeq B_1 \simeq A_1 \mintensor B_1$, while $S_1 :=
A_1 B_1$ is $A_1 \maxtensor B_1$. Since $A_1$ and $B_1$ are
non-classical, we can find an entangled state $\omega \in A_1
\maxtensor B_1$. If the composite were regular, we could apply the
entanglement-swapping protocol of Lemma 5 to pivot $\omega$ to an
entangled state on $B = B_1 \mintensor B_2$---which is absurd, as
the latter contains no
entangled states. \\

A similar disharmony obtains between the maximal and the usual
tensor products of quantum systems \cite{BFRW05}. Consider a
situation in which two quantum-mechanical systems, represented by
state spaces $A$ and $B$, are coupled by means of the maximal tensor
product to form $A \maxtensor B$. Suppose also that $A$ and $B$ are
themselves composite systems, say $A = A_1 \otimes A_2$ and $B =
B_1\otimes
  B_2$, where $\otimes$ is the usual quantum-mechanical tensor
product. Then an application of Lemma 5 shows that if $\omega$ is a
maximally entangled state on $A_2 \otimes B_2$ and $\rho \in A_1
\maxtensor B_1$ is what we might call an {\em ultra-entangled} state
of $A_1 \otimes B_1$---that is, a state of the maximal tensor
product not belonging to $A_1 B_1$---then conditional on a suitable
maximally entangled outcome for a measurement on $A$, one finds that
$\rho$ has apparently been teleported through $\omega$, and now
resides in $B_1 \otimes B_2$---which is absurd, as the latter is an
ordinary composite quantum system hosting no ultra-entangled states.
\\

\noindent{\bf 6. Conclusions and Prospectus} We have established
necessary and sufficient conditions for a composite of three
probabilistic models to admit a conclusive teleportation protocol.
We have also provided a class of examples illustrating that
deterministic teleportation can be supported by weakly self-dual
probabilistic models that are far from being either classical or
quantum-mechanical. Along the way, we have developed tools for
manipulating regular composites that are likely to be useful in any
systematic study of categories of probabilistic models, and
particularly categories equipped with more than a single tensor
product.

It remains an open problem to find non-trivial necessary and
sufficient conditions for a deterministic teleportation protocol to
exist. Theorem 3 is a step in this direction; however, one would like
a sharp criterion for the existence of a $G$-equivariant isomorphism
$A^{\ast} \simeq A$, where $G$ is a finite or, more generally, compact
group acting transitively on the extreme points of $\Omega_{A}$.

Looking further ahead, one would like to consider in detail the
categorical structure of probabilistic theories subject to precise
axioms governing remote evaluation, teleportation, etc., making
contact with the rapidly developing theory of information processing
in compact-closed categories \cite{Abramsky-Coecke, Baez04a,
Selinger2004a, Selinger2007a}.  \\

\noindent{\em Acknowledgements} Significant parts of this work were
done at the following conferences, retreats and workshops during 2007:
(i) New Directions in the Foundations of Physics, College Park, MD
(HB, JB, ML, AW); (ii) Philosophical and Formal Foundations of Modern
Physics, Les Treilles, (HB); (iii) Operational Theories as Foils to
Quantum Theory, Cambridge, supported by the Foundational Questions
Institute (FQXi) and SECOQC (HB, JB, ML, AW); (iv) Operational
Approaches to Quantum Theory, Paris (HB, AW).  We wish to thank the
organizers of these events, Jeffrey Bub and Rob Rynasiewicz, Tony
Short and Rob Spekkens, and Alexei Grinbaum, for the invaluable
opportunities they provided for us to work on this project.

At IQC, ML was supported in part by MITACS and ORDCF.  ML was
supported in part by grant RFP1-06-006 from FQXi.  Research at
Perimeter Institute for Theoretical Physics is supported in part by
the Government of Canada through NSERC and by the Province of Ontario
through MRI.  This work was also carried out partially under the
auspices of the US Department of Energy through the LDRD program at
LANL under Contract No. DE-AC52-06NA25396.

%{\bf References} %%Not all of these are actually used...

\newpage
\begin{appendix}

\noindent{\bf Appendix: proofs from section 3}
%\footnote{Here $\ostar$ has become $\odot$, since this
%has a large form...}\\

\noindent{\em Proof of Lemma 1} (a) Let $\mu \in (A^{J})_{+}$ be a
positive linear combination $\mu = \sum_p t_p (\omega_p)^{J}_{a^p}$
of reduced states, where for all $p$, $\omega_p \in A$ and $a^{p} =
(a^{p}_{i}) \in \Pi_{i \in I \setminus J} A_{i}^{\ast}$. Then for
any $b = (b_j) \in \Pi_{j \in J \setminus K}$, we have $\mu^{K}_{b}
= \sum_p t_p (\omega_{p})^{J}_{a^{p}}(b) = \sum_p
(\omega_{p})^{K}_{a^{p} \otimes b} \in A^K$. It follows that
$((A^J)^{K})_{+} \subseteq (A^{K})_{+}$. For the converse, let
$\omega \in A_{+}$: for any $a = (a_i) \in \Pi_{i \in I} A_{i}$, we
have $a = b \otimes c$ where $b = (b_j) \in \Pi_{j \in J \setminus
K} A_j$  and $c = (c_k) \in \Pi_{k \in K} A_k$. Thus,
$\omega^{K}_{a} = (\omega^{J}_{b})^{K}_{c} \in ((A^{J})^{K})_{+}$.

For (b), suppose $\omega \in A$ and $a = (a_i)$ in $\Pi_{i \not \in
I} A_{i}^{\ast}$. Pick any $c = (b_j) \in \Pi_{j \in J}
A_{j}^{\ast}$: we can set
\[\alpha = \omega^{J}_{a} \in A_J \ \text{and} \ \beta = \omega^{I -
J}_{b} \in A_{I - J}.\] If $A$ is regular, we then have $\alpha
\otimes \beta \in A$, whence,
\[\alpha = (\alpha \otimes \gamma)^{J}_{u_{I - J}}.\]
Part (c) follows from (a) and (b). %(details?)
$\Box$\\

\noindent{\em Proof of Proposition 1} Let $A = \bigodot_{i\in I}
A_i$. We first show that, for any set $J \subseteq I$, $A^{J} =
\bigodot_{j \in J} A_j)$. By assumption, we have
\[A \simeq (\bigodot_{j \in J} A_j) \odot (\bigodot_{k \in I
\setminus J} A_{k}) \geq (\bigodot_{j \in J} A_j) \mintensor
(\bigodot_{k \in I \setminus J} A_{k}).\] It follows that, for every
$\mu \in \bigodot_{j \in J} A_{j}$, and for any $\nu \in \bigodot_{k
\in I \setminus J} A_{k}$, $\mu \otimes \nu \in A$; hence,
\[\mu = (\mu \otimes \nu)^{J}_{\otimes_{k \in I \setminus J} u_k}
\in A_{J}.\] Thus, $\bigodot_{j \in J} A_j \leq A^{J}$.

For the reverse inclusion, note that we also have
\[(\bigodot_{j \in J} A_j) \odot (\bigodot_{k \in I \setminus J}
A_{k}) \leq (\bigodot_{j \in J} A_j) \maxtensor (\bigodot_{k \in I
\setminus J} A_{k});\] hence, for any $\omega \in A$ and any $f \in
(\bigodot_{k \in K} A^{k})^{\ast}$---in particular, for any $f =
(f_k)_{k \in I \setminus J}$---we have $\omega^{J}_{f} \in
\bigodot_{j \in J} A_j$. The rest of the proof now proceeds easily.
If $J_1,...,J_m$ is a partition of $I$, then we have $A =
\bigodot_{p = 1}^{m} (\bigodot_{j \in J_{p}} A_{p}) = \bigodot_{p
=1}^{m} A^{J_{p}}$. Since $\odot$ is a coupling, this last is a
composite of
$A^{J_{p}}$, $p = 1,...,m$. $\Box$\\

\end{appendix}


\begin{thebibliography}{References}
\bibitem{Abramsky-Coecke} S. Abramsky and B. Coecke, A categorical semantics of quantum
protocols, quant-ph/0402130v5 (2004, revised 2007)
\bibitem{Baez04a}
J.~Baez.
\newblock Quantum quandaries: a category-theoretic perspective.
\newblock {\tt quant-ph/0404040}, 2004.
\bibitem{Barrett} 2 J. Barrett, Information processing in general probabilistic
theories, Phys. Rev. A. {\bf 75} (2007) 032304-
\bibitem{BBLW06} H. Barnum, J. Barrett, M. Leifer and A. Wilce, Cloning and
Broadcasting in Generic Probabilistic Models, quant-ph/061129 (2006)
\bibitem{BBLW07} H. Barnum, J. Barrett, M. Leifer and A. Wilce, A
general no-cloning theorem, Phys. Rev. Lett. {\bf 99} 240501 (2007).
\bibitem{BFRW05}  H. Barnum, C. Fuchs, J. Renes and A. Wilce, Influence-free
states on coupled quantum-mechanical systems, quant-ph/0507108
(2005)
\bibitem{Bennett et al} C.H. Bennett, G. Brassard, C. Cr{\'e}peau, R. Jozsa, A. Peres and
W.K. Wootters, Teleporting an unknown quantum state via dual
classical and Einstein-Podolsky-Rosen channels, Physical Review
Letters, Vol. 70 (1993), 1895 – 1899.
\bibitem{DaviesLewis} E. B. Davies and J. T. Lewis, An operational approach to
quantum probability, Comm. Math. Phys. {\bf 17} (1970) 239-260
\bibitem{Edwards} C. M. Edwards, The operational approach to quantum probability I, Comm. Math.
Phys. {\bf 17} (1971), 207-230.
\bibitem{Ellis} A. J. Ellis, Linear operators in partially ordered normed vector spaces, J. London Math. Soc. {\bf 41} (1966)
323-332.
\bibitem{HolevoJMP} A. Holevo,
Radon-Nikodym derivatives of quantum instruments J. Math. Phys. {\bf
39} (1998) 1373-
%\bibitem{Horne} J. G. Horne, ``On the automorphism group of a cone,'' Lin. Alg. Appl. {\bf 21}
%(1978) 111--121.
\bibitem{Hardy} L. Hardy, A framework for probabilistic theories with non-fixed causal structure, J. Phys. A. {\bf 40} (2007) 3081
\bibitem{Hardy2} L. Hardy, Disentangling nonlocality and teleportation quant-ph/9906123 (1999)
\bibitem{Klay} M. Kl\"{a}y, D. J. Foulis, and C. H. Randall, Tensor products and
probability weights, Int. J. Theor. Phys. {\bf 26} (1987), 199-219.
\bibitem{Koecher} Koecher, Die geood\"{a}tischen von Positivita\"{a}tsbereichen, Math. Annalen {\bf 135} (1958)
192-202.
\bibitem{Ludwig} G. Ludwig, {\em An Axiomatic Basis of Quantum
Mechanics 1, 2}, Springer-Verlag, 1985, 1987.
\bibitem{Mackey} G. Mackey, {\em Mathematical Foundations of Quantum
Mechanics}, Benjamin, 1963.
\bibitem{Namioka-Phelps} I. Namioka and R. Phelps, Tensor products of compact convex
sets, Pacific J. Math. {\bf 9} (1969), 469-480.
\bibitem{Selinger2004a}
P.~Selinger.
\newblock Towards a semantics for higher-order quantum computation.
\newblock In {\em \em Proceedings of the 2nd International Workshop on Quantum
  Programming Languages, Turku Finland}, pages 127--143. Turku Center for
  Computer Science, 2004.
\newblock Publication No. 33.
\bibitem{Selinger2007a}
P.~Selinger.
\newblock Dagger compact closed categories.
\newblock {\em Electronic Notes in Theoretical Computer Science}, 170:139--163,
  2007.
\newblock Proceedings of the 3rd International Workshop on Quantum Programming
  Languages (QPL 2005), Chicago.
\bibitem{Spekkens} R. Spekkens, Evidence for the epistemic view of
quantum states: a toy theory, Phys. Rev. A. {\bf 75} (2007) 032110
\bibitem{Wittstock} G. Wittstock, Ordered normed tensor products, in H. Neumann and
H. Hartkamper (eds.), Foundations of quantum mechanics and ordered
linear spaces, Springer Lecture Notes in Physics, 1974.
\bibitem{Vinberg} E. B. Vinberg, Homogeneous cones, Dokl. Acad. Nauk.  SSSR {\bf 141} (1960)
270-273; English trans. Soviet Math. Dokl. {\bf 2} (1961) 1416-1619.
\\
%(Howard, can you supply the reference?)



\end{thebibliography}
\end{document}